\documentclass[%
prl%
,preprint%
 ,secnumarabic%
,amssymb, amsmath,nobibnotes, aps, prl]{revtex4}
\usepackage{epsfig}%
\usepackage{graphicx}%
\expandafter\ifx\csname package@font\endcsname\relax\else
 \expandafter\expandafter
 \expandafter\usepackage
 \expandafter\expandafter
 \expandafter{\csname package@font\endcsname}%
\fi

\begin{document}

\title{Test of unparticle long range forces from perihelion precession of Mercury}%

\author{Suratna Das ,   Subhendra  Mohanty and Kumar Rao }%
\address{Physical Research Laboratory, Ahmedabad 380009,
India.}
\def\be{\begin{equation}}
\def\ee{\end{equation}}
\def\al{\alpha}
\def\bea{\begin{eqnarray}}
\def\eea{\end{eqnarray}}

\begin{abstract}
Unparticle exchange gives rise to long range forces which deviate
from the inverse square law due to non-canonical dimension of
unparticles. It is well known that a potential of the form
$r^{-n}$ where $n$ is not equal to one gives rise to a precession
in the perihelion of planetary orbits. We calculate the
constraints on unparticle couplings with baryons and leptons from
the observations of perihelion advance of Mercury orbit.

PACS : 98.80.Cq, 11.15.Tk, 11.25.Hf
\end{abstract}

\maketitle
Recently a new class of particles with dimensions different from
their canonical scaling dimensions have been proposed to exist in
the effective low energy theory \cite{georgi}.  One assumes that
an ultraviolet theory has a IR fixed point at some scale
$\Lambda_u$ where the fields become conformal invariant. The
effective coupling of the ultraviolet theory operators $O_{UV}$ of
dimension $d_{uv}$ with the standard model operators $O_{SM}$ of dimension
$n$ are suppressed by a heavy mass scale $M_u$ and can be written
as \be \frac{1}{M_u^{d_{uv}+n-4}}\,O_{UV} O_{SM}, \ee where
$d_{uv}$ is the canonical dimension of the operator $O_{UV}$.
Below the scale $\Lambda_u$ (conventionally assumed as 1 TeV), the
fields of the UV theory become scale invariant and by dimensional
transmutation acquire a dimension $d_u$ which is different from
their canonical dimension. These conformally coupled unparticle
operators $O_U$ will couple to the standard model operators as \be
\left(\frac{\Lambda_u}{M_u} \right)^{d_{uv}+n-4} \frac{
1}{\Lambda_u^{d_u+n-4}}\,O_U \, O_{SM}. \ee

It has been pointed out \cite{liao} that the exchange of scalar (pseudoscalar) unparticles can give rise to spin independent (spin-dependent) long range forces.
Long range forces from vectors and axial-vectors have been studied in \cite{liao,desh}. Tensor unparticles can couple to the energy momentum tensor and mimic gravity as pointed out in \cite{nath}.
Unparticle exchange  gives rise to long range forces
which deviate from the usual inverse square law for massless
particles due to the anomalous scaling of the unparticle propagator.
In \cite{nath, desh} bounds have been put on the unparticle
couplings from long range force experiments \cite{long}.

 It is
well known that a deviation from the Newtonian inverse square
gravity will result in unclosed  orbits which results in a shift
in perihelion of planetary orbits. Since exchange of massless
unparticles gives rise to long range forces which deviate from the
inverse square law we expect an additional contribution to the
perihelion shift of planets in addition to that caused by general
relativity.
 In this paper we consider the effect on
the perihelion shift of Mercury due to the coupling of tensor and
vector unparticles  to SM particles. The perihelion shift due to
general relativistic effects has been measured to $0.3 \%$ level and
thus provides tight constraints on additional long range forces \cite{sereno}.
We find that this gives  more stringent bounds on unparticle
couplings compared to the one from fifth force search experiments
at solar system distances \cite{long}.

Some consequences of unparticles in astrophysical phenomena has been
explored in \cite{wyler,hooman,raffelt,das}. There has also been a
large amount of work on the theory and phenomenology of
unparticles \cite{all}.

\section{Ungravity from tensor unparticles}
We take the gravitational coupling of the tensor unparticle
(ungravitons \cite{nath}) to the stress-energy tensor $T_{\mu \nu}$ to be of
the form
\begin{equation}
\kappa_{*}\frac{1}{\Lambda_u^{d_u-1}}\sqrt{g}T^{\mu\nu}O^{U}_{\mu\nu},
\end{equation}
where
$\kappa_{*}=\frac{1}{\Lambda_u}\left(\frac{\Lambda_u}{M_u}\right)^{d_{uv}}$.
We impose the gauge symmetry as in the case of gravity, \bea x_\mu
&\rightarrow& x_\mu + \epsilon_\mu \\ O_{\mu \nu}^U &\rightarrow&
O_{\mu \nu}^U+\frac{\Lambda_u^{d_u-1}}{\kappa_*}\left(\partial_\mu
\epsilon _\nu + \partial_\nu \epsilon _\mu \right),
\label{gaugesymmetry}
\eea 
which ensures that the ungraviton remains massless below the
scale $\Lambda_u$. The massless ungraviton results in long range
forces which can be probed at solar system length scales.

The ungraviton propagators are \cite{nath}
\begin{equation}
\Delta^{\mu\nu\alpha\beta}(P)=B_{d_u}P^{\mu\nu\alpha\beta}(-P^2)^{d_u-2},
\label{propagator}
\end{equation}
where the normalization factor $B_{d_u}$ is
\begin{equation}
B_{d_u}\equiv-\left(\frac{8\pi^{\frac32}}{\left(2\pi\right)^{2d_u}}\right)\frac{\Gamma\left(2-d_u\right)\Gamma\left(d_u+\frac12\right)}{\Gamma\left(2d_u\right)},
\end{equation}
and $P^{\mu\nu\alpha\beta}$ is the projection operator of the form
\begin{equation}
P^{\mu\nu\alpha\beta}(P)\equiv
\frac12(P^{\mu\alpha}P^{\nu\beta}+P^{\mu\beta}P^{\nu\alpha}-\alpha
P^{\mu\nu}P^{\alpha\beta}),
\end{equation}
where $P^{\mu\nu}=\left(-\eta^{\mu\nu}+\frac{P^\mu
P^\nu}{P^2}\right)$.  For massless ungravitons, obeying the gauge
condition of Eq~(\ref{gaugesymmetry}), $\alpha =1$.

The ungravitational potential is obtained by taking the Fourier
transform of the propagator $\Delta^{\mu\nu\alpha\beta}$ in the static
limit $\left(P^0=0\right)$ : 
\begin{equation} 
V_u(r)=\frac{\kappa_*^2}{\Lambda_u^{2d_u-2}} \int \frac{d^3{\bf P}}{(2
\pi)^3} T_{\mu \nu} \Delta^{\mu \nu \alpha \beta}(P^0=0)T_{\alpha
\beta} e^{i{\bf P}\cdot {\bf x}},
\end{equation} 
where $|{\bf x}|=r$. Evaluating the integral gives
\begin{eqnarray}
V_u(r)&=&-m_1m_2\left(\frac{\kappa_{*}^2}{\Lambda_u^{2d_u-2}}\right)\left(\frac2{\pi^{2d_u-1}}\right)\frac{\Gamma\left(d_u+\frac12\right)\Gamma\left(d_u-\frac12\right)}{\Gamma\left(2d_u\right)}\left(\frac1{r^{2d_u-1}}\right) \nonumber \\
&=& -\frac{G_u m_1 m_2}{r^{2d_u-1}},
\end{eqnarray}
where $G_u$ is defined to be
\begin{equation}
G_u\equiv\frac{\kappa_{*}^2}{\Lambda_u^{2d_u-2}}C(d_u),
\label{eq:gu}
\end{equation}
and $C(d_u)$ is
\begin{equation}
C(d_u)\equiv\left(\frac2{\pi^{2d_u-1}}\right)\frac{\Gamma\left(d_u+\frac12\right)\Gamma\left(d_u-\frac12\right)}{\Gamma\left(2d_u\right)}.
\end{equation}
We notice that if the anomalous dimension $(d_u)$ of $O_{\mu \nu}$ is
not equal to 1 there are deviations from the inverse square law. So
for $d_u\neq 1$ the total potential will be of the form :
\begin{eqnarray}
V(r)&=&-\frac{Gm_1m_2}{r}-\frac{G_um_1m_2}{r^{2d_u-1}}. \nonumber \\
&=&-\frac{Gm_1m_2}{r}\left[1+\frac{1}{G\Lambda^2_u}\left(\frac{\Lambda_u}{M_u}\right)^{2d_{uv}}\frac{C(d_u)}{\Lambda_u^{2d_u-2}}\frac1{r^{2d_u-2}}\right].
\label{eq:vr}
\end{eqnarray}
We will consider the case $d_u>1$ as $d_u<1$ will lead to forces which
fall off slower than gravity and can be easily ruled out from fifth
force experiments \cite{long}.

\subsection{Perihelion precession of mercury orbit}

In polar co-ordinates $(r,\theta)$, the equation of motion of a
planet's orbit around the Sun is
\begin{equation}
\ddot{r}-r\dot{\theta}^2+\frac{V^{\prime}(r)}{m_p}=0,
\label{eq:eom}
\end{equation}
where $m_p$ is the mass of the planet and  $\dot{}$ and $^\prime$
represent derivatives with respect to time $t$ and distance $r$
respectively. The  angular momentum of the planet
$l=m_pr^2\dot{\theta}$ is a constant of motion.

Changing variables to $u(\theta)=\frac{1}{r\left(\theta\right)}$,
Eq~(\ref{eq:eom}) can be written as \be u^{\prime\prime}+u
=\alpha+\beta u^{2d_u-2}.\label{eom} \ee Here $^\prime$ represents
derivative with respect to $\theta$ and
$\alpha\equiv\frac{Mm_p^2G}{l^2}$ and $\beta\equiv\frac{Mm_p^2
G_u(2d_u-1)}{l^2}$, where $M$ is the mass of the Sun. This is an
inhomogeneous second order ordinary differential equation.
Assuming the deviation from the inverse square law to be very
small, we have $\beta<<\alpha$. So Eq~(\ref{eom}) can be solved using
a perturbation expantion in $\beta$. To first order in $\beta$ we
assume the form of the solution to be
\begin{equation}
u(\theta)=u_0(\theta)+\beta u_1(\theta), \label{equ}
\end{equation}
where $u_0$ is the solution of the ODE
\begin{equation}
u_0^{\prime\prime}+u_0=\alpha, \label{equ0}
\end{equation}
and $u_1$ is the particular solution of the inhomogeneous
equation
\begin{equation}
u_1^{\prime\prime}+u_1=u_0^{2d_u -2}. \label{equ1}
\end{equation}
The solution of Eq~(\ref{equ0}) is
\begin{equation}
u_0=\frac{1-e\cos\left(\theta\right)}{a(1-e^2)}, \label{uo}
\end{equation}
where $a$ is the semi-major axis of the elliptical orbit of the
planet, given by \be a=\frac{l^2}{M m_p^{2} G (1-e^2)} \label{u0} \ee
and $e$ is the eccentricity of the orbit. As the eccentricity of
Mercury's orbit is very small we keep terms only upto ${\cal O}(e)$ and
neglect the higher order terms in Eq~(\ref{uo}). Using the above form of $u_0$,
$u_1(\theta)$ obeys the equation
\begin{equation}
u_1^{\prime\prime}+u_1=\frac1{a^{2d_u-2}}-\frac{\left(2d_u-2\right)e\cos\left(\theta\right)}{a^{2d_u-2}}.
\end{equation}
This has the particular solution
\begin{equation}
u_1=\frac1{a^{2d_u-2}}-\frac{\left(d_u-1\right)e}{a^{2d_u-2}}\theta
\sin(\theta).
\end{equation}
Thus, from Eq~(\ref{equ}), the trajectory of the planet to order
$\beta$ is given by
\begin{equation}
u
=\frac1a+\beta\frac1{a^{2d_u-2}}-\frac{e}{a}\left[\cos(\theta)+\frac{\beta\left(d_u-1\right)}{a^{2d_u-3}}\theta\sin(\theta)\right].
\label{eq:usol}
\end{equation}
For small $\beta$, Eq~(\ref{eq:usol}) can be written as
\begin{equation}
u\approx \frac1a+\beta\frac1{a^{2d_u-2}}-\frac{e}{a}\left[\cos
\left(\theta-\frac{\beta\left(d_u-1\right)}{a^{2d_u-3}}\theta\right)\right].
\end{equation}
For one complete rotation with a perihelion shift the condition is
\begin{equation}
\theta\left(1-\frac{\beta\left(d_u-1\right)}{a^{2d_u-3}}\right)=2\pi,
\end{equation}
which gives
\begin{equation}
\theta\approx2\pi\left(1+\frac{\beta\left(d_u-1\right)}{a^{2d_u-3}}\right),
\label{eq:theta}
\end{equation}
keeping only terms linear in $\beta$. So the perihelion shift
induced by ungraviton couplings is given by
\begin{eqnarray}
\delta \theta&=&
2\pi\left(\frac{\beta\left(d_u-1\right)}{a^{2d_u-3}}\right) \\
\nonumber &=& (d_u-1)(2d_u-1)C(d_u)
\frac{2\pi}{G\Lambda^2_u}\left(\frac{\Lambda_u}{M_u}\right)^{2d_{uv}}\frac{1}{\Lambda_u^{2d_u-2}}\frac1{a^{2d_u-2}}.
\label{thetaUn}
\end{eqnarray}
As expected, the perihelion shift vanishes for $d_u=1$, as it should
since it corresponds to the usual inverse square law case (with a
different gravitational constant). Comparing the expression for the
unparticle potential Eq~(\ref{eq:vr}) (for $r=a$) with the expression
for perihelion advance we see that they are related as \be \delta
\theta \simeq (d_u-1)(2d_u-1) 2 \pi \frac{V_u}{V_N}, \ee where $V_u$
is the unparticle exchange potential and $V_N$ is the Newtonian
potential. 
The constraint on the ungravity couplings derived from mercury
perihelion are more stringent than that from fifth force measurement
by testing deviation from Kepler's Law at planetary distances
\cite{nature},\cite{prl}. However at milimeter scales there are stringent tests of deviations of Newton's Law as has been noted in \cite{nath},\cite{desh}.

The observed precession of perihelion of mercury is $43.13 \pm 0.14$
arcsec/century \cite{mercury} and the prediction from general
relativity is 42.98 arcsec/century. This means that at 2-$\sigma$ the
unparticle contribution is $-0.13 < \delta \theta < 0.43$. We derive a
limit on unparticle coupling by demanding that the unparticle
contribution does not exceed the discrepency between measurement and
GR. From the 2-$\sigma$ upper bound on the possible contribution from
unparticle given by Eq~(\ref{thetaUn}) we get the limit
\begin{eqnarray}
 (d_u-1)(2d_u-1)C(d_u)\frac{2\pi}{G\Lambda^2_u}\left(\frac{\Lambda_u}{M_u}\right)^{2d_{uv}}\frac{1}{(a \Lambda_u)^{2d_u-2}}\left(\frac{{\rm century}}{T}\right)<0.43\,\,\textrm{arcsec}
\label{thetaUn1}
\end{eqnarray}
per century, where $T=87.96$ days is the orbital time period of Mercury. 

\begin{figure}[h]
\centering
\includegraphics[width=0.75\textwidth]{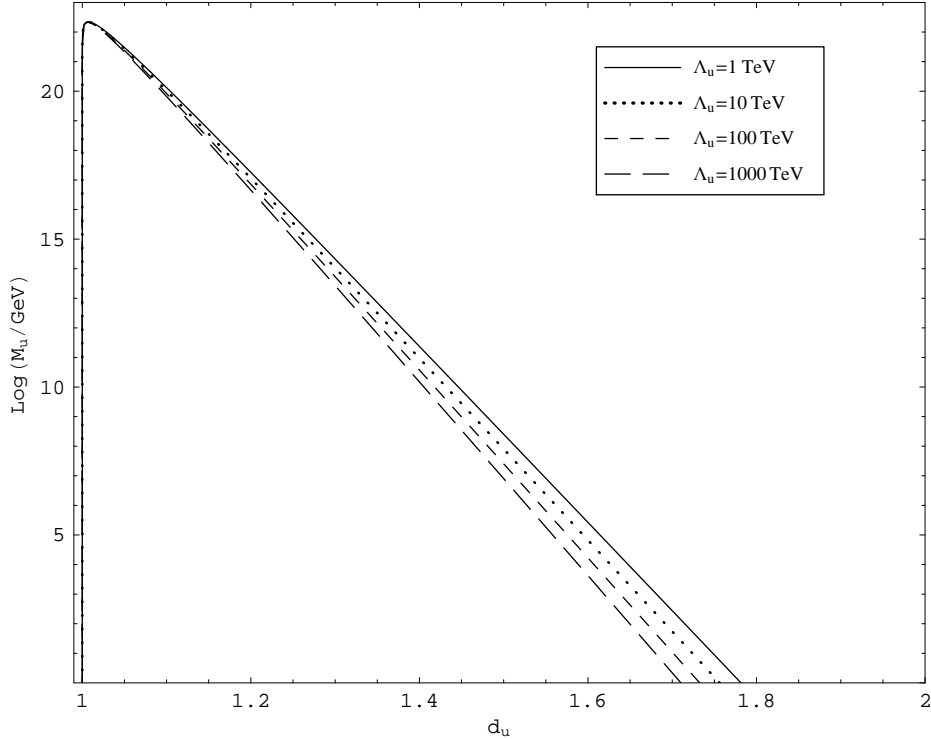}
\caption{Regions above the curves represent the allowed values of
{$\log\left(\frac{M_u}{{\rm GeV}}\right)$ and $d_u$ from observations of
Mercury orbit.  }}
\label{fig1}
\end{figure}

In Fig~(\ref{fig1}) we plot $\log\left(\frac{M_u}{{\rm GeV}}\right)$ vs
$d_u$ which gives the tensor unparticle contribution of $0.43$
arcsec/century to the perihelion advance of mercury. We have taken
$d_{uv}=1$ and the values of $\Lambda_u$ from 1 TeV to 1000 TeV. The
areas above the curves represent the allowed regions for $M_u$ and
$d_u$ at 2-$\sigma$ for different values of $\Lambda_u$ .

\section{Long range force from vector unparticles}
Now we consider long range forces resulting from the coupling of
vector unparticles \cite{liao,desh} to baryonic and leptonic currents. The
effective coupling is of the form
\begin{equation}
\frac{\lambda}{\Lambda_u^{d_u-1}}J^\mu O_{\mu}^{U},
\end{equation}
where $J_{\mu}$ is the baryonic or leptonic current. As in the
tensor case, we assume that the unparticle operator $O^U$ and the
fermion fields $\Psi$ obey a gauge symmetry
\begin{eqnarray}
\Psi &\to& \exp[i \alpha]\Psi \nonumber \\
O_{\mu}^{U}&\to& O_{\mu}^{U}+ \frac{\Lambda_u^{d_u -1}}{\lambda}\partial_{\mu}
\alpha.
\end{eqnarray}
As a result of this $U(1)$ gauge symmetry the vector unparticle
remains massless below the scale $\Lambda_u$. The gauge unparticle
propagator is
\begin{equation}
\Delta^{\mu \nu}= A_{d_u}P^{\mu \nu} (-p^2)^{d_u -2}, \label{vecprop}
\end{equation}
where 
\be
A_{d_u}\equiv\frac{16\pi^{\frac52}}{\left(2\pi\right)^{2d_u}}\frac{\Gamma\left(d_u+\frac12\right)}{\Gamma\left(d_u-1\right)\Gamma\left(2d_u\right)},
\ee
and 
\be P^{\mu \nu}(p)=\eta^{\mu \nu}-\frac{p^{\mu} p^{\nu}}{p^2} .\ee
As usual, we get the unparticle exchange potential by taking the
Fourier transform of the propagator given in Eq~(\ref{vecprop}) in the static
limit. This gives
\begin{eqnarray}
V_u(r)&=&\frac{1}{2\pi^{2d_u}}\frac{\lambda^2}{\Lambda_u^{2d_u-2}}\frac{\Gamma\left(d_u+\frac12\right)\Gamma\left(d_u-\frac12\right)}{\Gamma\left(2d_u\right)}\frac{N_1N_2}{r^{2d_u-1}} \nonumber \\
&=&\frac{C^{\prime}\left(d_u\right)\lambda^2N_1N_2}{r^{2d_u-1}},
\end{eqnarray}
where
\begin{equation}
C^{\prime}(d_u)\equiv\frac1{2\pi^{2d_u}}\frac1{\Lambda_u^{2d_u-2}}\frac{\Gamma\left(d_u+\frac12\right)\Gamma\left(d_u-\frac12\right)}{\Gamma\left(2d_u\right)},
\end{equation}
is a constant and $N_1$ and $N_2$ are the total number of baryons
($N_i=\frac{M_i}{m_n}$,where $M_i$ is the mass of the sun or the
planet and $m_n$ is the nucleon mass) in the Sun and the planet. Hence
the total potential is
\begin{eqnarray}
V(r)&=&V_{N}(r)+V_u(r) \nonumber \\
&=&-\frac{Gm_1m_2}{r}\left[1-\frac{C^{\prime}\left(d_u\right)\lambda^2N_1N_2}{Gm_1m_2}\frac1{r^{2d_u-2}}\right].
\end{eqnarray}
By following the same methodology as in the tensor case we find
the perihelion shift to be
\begin{equation}
\delta\theta=-2\pi(d_u-1)(2d_u-1)\frac{C^{\prime}\left(d_u\right)\lambda^2N_1N_2}{Gm_1m_2}\frac1{a^{2d_u-2}}.
\end{equation}
\begin{figure}[h]
\centering
\includegraphics[width=0.75\textwidth]{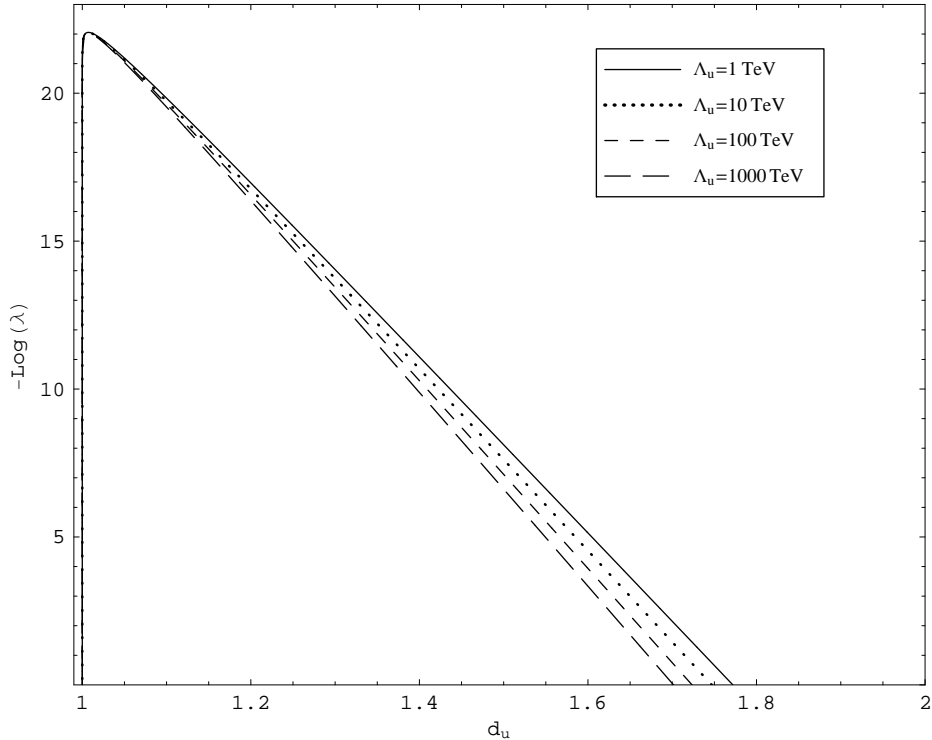}
\caption{Region above the curve represents the allowed values of
$-\log(\lambda)$ and $d_u$ from observations of Mercury orbit. }
\label{fig2}
\end{figure}
 Vector unparticle exchange would cause a retardation in the
 perihelion of mercury orbit ( $\delta \theta <0$) due to the fact
 that the force is repulsive. At 1-$\sigma$ the discrepancy between
 theory and experiment is still positive ($0.01 < \delta \theta <
 0.29$) which means that the vector unparticle force can be ruled out
 at 1-$\sigma$. At $2-\sigma$ the allowed range for a unparticle
 vector contribution is $-0.13 < \delta \theta < 0.43 $.  The maximum
 value of this retardation allowed from observations \cite{mercury}
 and the prediction of general relativity is 0.13 arcsec/century at
 2-$\sigma$.  This puts an upper bound on the vector unparticle
 couplings
\begin{equation}
 2\pi(d_u-1)(2d_u-1)\frac{C^{\prime}\left(d_u\right)\lambda^2N_1N_2}{Gm_1m_2}\frac1{a^{2d_u-2}}\left(\frac{{\rm century}}{T}\right) < 0.13 \,\,\textrm{arcsec}
\end{equation}
per century, where $T=87.96$ days is the orbital time period of Mercury as stated
before.

In Fig~(\ref{fig2}) we show $-\log\left(\lambda\right)$ vs. $d_u$ plot
taking $\delta \theta=0.13$ arcsec/century. We have taken the values
of $\Lambda_u$ from 1 TeV to 1000 TeV. The areas above the curves
represent the allowed values of $\lambda$ and $d_u$ at 2-$\sigma$
experimental error for different $\Lambda_u$  after accounting for
the contribution to perihilion shift from general relativity.

\section{Conclusions}
 There are several bounds on unparticle couplings to standard model
particles from collider experiments \cite{all} from the anomalous
missing energy spectrum. There are also bounds on such couplings from
the cooling rates of supernova and stars \cite{wyler, hooman, das,
raffelt}. If the conformal invariance of unparticles remains unbroken
then these particles can give rise to extra long range forces
\cite{nath, desh} which can be constrained from fifth force
experiments \cite{long}. In this paper we have considered unparticle
gauge bosons of spin-1 and spin-2. The gauge symmetry ensures that the
unparticles remain massless. The main characteristic feature of
unparticle long range force which we apply in this paper is a
deviation from the inverse square law which leads to a perihelion
shift in planetary orbits. The constraints from perihelion shift are
more stringent than the constraints from the deviation from the
inverse square law at the scale of solar system distances
\cite{nature,prl}. However at milimeter scales there are stringent
tests of deviations of Newton's Law as has been noted in
\cite{nath,desh}. Compairing our bounds on vector and tensor
unparticle couplings with that of \cite{nath} and \cite{desh} we find
our bounds based on perihelion precession are more stringent when
$d_u\lesssim 1.4$.



\end{document}